# Ferroelectric amplitude switching and continuous memory

Gye-Hyeon Kim [1]†, Tae Hyun Jung [2]†, Seungjoon Sun [1]†, Jung Kyu Lee [2], Jaewoo Han [1], P. Karuna Kumari [3], Jin-Hyun Choi [1], Hansol Lee [1], Tae Heon Kim [4,5], Yoon Seok Oh [1], Seung Chul Chae [6], Se Young Park [3], Sang Mo Yang [2]*, Changhee Sohn [1]*

[1]Department of Physics, Ulsan National Institute of Science and Technology, Ulsan 44919, Republic of Korea.

[2]Department of Physics, Sogang University, Seoul 04107, Republic of Korea.

[3]Department of Physics, Soongsil University, Seoul 06978, Republic of Korea.

[4]Electronic and Hybrid Materials Research Center, Korea Institute of Science and Technology, Seoul 02792, Republic of Korea.

[5]Division of Nanoscience and Technology, KIST School, Korea National University of Science and Technology, Seoul 02792, Republic of Korea.

[6]Department of Physics Education and Center for Education Research, Seoul National University, Seoul 08826, Republic of Korea.

†These authors contributed equally to this work: Gye-Hyeon Kim, Tae Hyun Jung, Seungjoon Sun,

*Corresponding authors. Email: Sang Mo Yang, smyang@sogang.ac.kr and Changhee Sohn, chsohn@unist.ac.kr

**Abstract:** Although ferroelectric systems inherently exhibit binary switching behavior, recent advances in analog memory device have spurred growing interest in achieving continuous memory states. In this work, we demonstrate ferroelectric amplitude switching at the mesoscopic scale in compositionally graded $Ba_{1-x}Sr_xTiO_3$ heterostructures, enabling continuous modulation of polarization magnitude without altering its direction, which we defined as amplitude switching. Using switching current measurement, piezoresponse force microscopy and Landau-Ginzburg-Devonshire simulations, we reveal that compositionally graded ferroelectric heterostructure can possess amplitude switching behavior through a double well potential with flattened minima. This behavior supports stable, continuous polarization states and establishes a new platform for analog memory applications. These findings introduce amplitude switching as a new dynamic of the order parameter, paving the way for energy-efficient and reliable analog memory systems.

**Introduction**

Surprisingly large number of physical systems, ranging from our universe to fridge magnets, often crystallize into symmetry-broken phases with finite order parameters (*1*). This ubiquitous behavior originates from the generic shape of the free energy landscape below the critical temperature, as described within the Ginzburg-Landau paradigm: a double-well potential for discrete symmetry breaking, or a Mexican hat potential for continuous symmetry. In ferroelectrics, for instance, the effective free energy exhibits symmetric double-well potential with respect to the order parameter, electric polarization, as schematically shown in Fig. 1A (*2-4*). As a result, the ground state can reside in one of two stable states (say, "up" or "down" polarization), separated by an energy barrier, with finite electric polarization even without external electric field. Exploiting such robust two-state physics is the foundation of binary computing: one classical bit of information is reliably stored by locating a system in one well or the other. Indeed, nonvolatile memory technologies using broken-symmetry phases capitalize on this profound thermodynamic preference for binary order (*5-7*).

This symmetry breaking that gifts us robust binary states also led to a seemingly trivial consequence: modulation of an order parameter's amplitude is fundamentally prohibited at low energy. Once an order parameter (e.g. magnetization, polarization, charge-density wave/superconductivity order parameters) develops, its magnitude is locked to a specific value in each stable phase, and changing its amplitude invokes an energy cost akin to exciting a massive amplitude mode (*8-12*), such as Higgs modes in gauge symmetry breaking (*13-16*). In the case of ferroelectrics, while the direction of polarization between "up" and "down" is readily switched by an external electric field (referred to as "phase switching" hereinafter), it is inherently unstable to partially increase the magnitude of polarization (amplitude switching): the system naturally tends to relax toward the nearest global minimum state (*17*). Yet, if such amplitude modulation were made possible, it could unlock a fundamentally new operating mode for order parameters—one that enables robust, continuous states and offers breakthrough potential for analog memory and neuromorphic computing.

Here, we present the realization of ferroelectric amplitude switching at the mesoscopic scale in a compositionally graded $Ba_{1-x}Sr_xTiO_3$ (CG-BSTO) heterostructure by engineering layer-dependent doping ratios (*18, 19*). In a conventional ferroelectric such as $BaTiO_3$ (BTO) thin film, the $Ti^{4+}$ ion within the $TiO_6$ octahedron dictates the direction of electric polarization, yielding a binary "up" or "down" state. This exemplifies the phase switching, where only two stable minimum states (0 or 1) exist, in a double-well potential (Fig. 1A) (*2, 3*). However, in CG-BSTO heterostructures with different Sr doping ratios across layers, the energy landscape at the mesoscopic scale becomes flattened near its minima, enabling further increase of polarization after full phase switching (Fig. 1B). This incrementally increasing in the magnitude of the polarization state corresponds to the aforementioned amplitude switching, providing a route toward continuous memory systems. Such continuous polarization states are found to be nonvolatile and therefore, may provide a pathway toward analog memory applications.

**Result**

**Material design for compositionally graded ferroelectric heterostructure**

We synthesized the BTO thin film and CG-BSTO heterostructure with top and bottom SrRuO$_3$ (SRO) electrodes using pulsed laser deposition and confirm its high crystallinity and well-controlled layer-dependent doping through X-ray diffraction, scanning transmission electron microscopy (STEM), and time-of-flight secondary ion mass spectrometry (TOF-SIMS) (Fig. 1C, Materials and Methods 1-4, and fig. S1, S2). As shown in Fig. 1C, a direct comparison between the TEM image and TOF-SIMS analysis reveals a well-defined doping gradient, where the Sr concentration gradually increases from the top surface toward the bottom SRO electrode, while the Ba concentration decreases accordingly within the same layers. Based on the relative intensity of Ba and Sr signals in TOF-SIMS between the top and bottom parts of heterostructure, the Sr doping ratio of our CG-BSTO heterostructures almost linearly change from 0 (top surface) to 0.26 (bottom interfaces) within 96 nm sample thickness. High-resolution TEM imaging confirms the epitaxial relationship and atomically sharp interfaces within the heterostructure. The ferroelectric properties of both BTO and the CG-BSTO heterostructure were verified through polarization-voltage (*P-V*), current-voltage (*I-V*), capacitance-voltage (*C-V*), leakage current measurements, and polarization relaxation measurement (fig. S3 – S5). Both BTO and CG-BSTO show 1) square-like *P-V* hysteresis loops with small difference between the saturated polarization and remanent polarization across the 1-5 kHz range, 2) low leakage current density, and 3) excellent stability of the polarization state over time. The set of experiments consistently confirms robust ferroelectricity in both heterostructures, with no indication of relaxor-like behavior (*20*).

**Experimental observation of amplitude switching in ferroelectric polarization in compositionally graded ferroelectric heterostructure**

To investigate the polarization switching characteristics, we performed switching current measurements on BTO and CG-BSTO capacitor structures (Materials and Methods 5) (*21-24*). Fig. 2A exhibits the time-dependent normalized switched polarization, $\Delta P(t)/2P_r$ (open squares) of BTO. The switching kinetics obtained in BTO show an abrupt increase in $\Delta P(t)/2P_r$ up to around 4 μs, followed by saturation. This trend matches well with the Kolmogorov-Avrami-Ishibashi (KAI) model (gray dashed line), which relies on classical statistical theory of nucleation and unrestricted domain growth (*21, 25*). However, the polarization switching behavior observed in the CG-BSTO heterostructure is markedly different from that in the BTO film. As shown in Fig. 2B, while $\Delta P(t)/2P_r$ (open squares) in CG-BSTO increases drastically in the initial stage of switching (within 5 μs), similar to BTO, it continues to increase over a much longer timescale (up to 10 s). This anomalous polarization switching behavior in the CG-BSTO heterostructure cannot be explained by either the KAI model or other polarization switching models such as the nucleation-limited-switching model (*26*). We speculate that two temporally distinct polarization switching processes coexist in CG-BSTO: a rapid switching process at short timescales, and a relatively slower switching process occurring over long timescales. This anomalous switching behavior is consistently observed under various pulse conditions and regardless of the polarization direction, suggesting that it is an intrinsic characteristic of the CG-BSTO heterostructure, likely arising from its compositional gradient (fig. S6 - S8).

To gain direct nanoscale insight into the anomalous switching behavior observed in CG-BSTO in macroscopic current-based measurements, we performed time-resolved piezoresponse force microscopy (PFM) imaging (*27*). Specifically, we applied 1.5 V writing pulses with incrementally increasing durations to oppositely pre-poled capacitors via the SRO top electrode and acquired time-dependent PFM images (Materials and Methods 6) (*28-30*). As shown in Fig. 2C, the PFM amplitude and phase images of BTO illustrate the conventional domain switching process, including nucleation, growth of oppositely polarized domains, and their coalescence. Homogeneous PFM phase contrast and saturated amplitude signals are observed around 2 μs. Extending the writing pulse duration beyond this point (e.g., 10 μs or more) leads to no significant change in the PFM amplitude. These observations are consistent with the conventional KAI model. To ensure the reliability of the acquired PFM data, we extracted the normalized out-of-plane piezoresponse, *q* value (solid triangles in Fig. 2A) from the PFM images (Materials and Methods 6), and compared them with the switching current data. The normalized *q* values show good agreement with the $\Delta P(t)/2P_r$ values obtained from switching current measurements, validating the reliability of our PFM images (*24*). Therefore, we conclude that BTO exhibits only ferroelectric phase switching, where the polarization state shifts between two opposite directions without involving any significant modulation in magnitude.

In a similar manner, we investigated nanoscale domain switching behavior in CG-BSTO. As shown in Fig 2D, while the PFM phase in CG-BSTO becomes homogeneous (brown) within the microsecond scale—indicating the completion of domain reversal into the opposite polarization state—the PFM amplitude continues to increase over a much longer timescale, extending up to 10 s. This suggests that the polarization magnitude continues to increase even after the phase switching is complete. Such behavior is not observed in conventional ferroelectric systems, such as BTO, that undergo only the ferroelectric phase switching. Therefore, this direct observation goes beyond the framework of conventional switching mechanisms and confirms the emergence of ferroelectric amplitude switching. The normalized *q* values of CG-BSTO plotted in Fig. 2B (solid triangles) also show good agreement with $\Delta P(t)/2P_r$, validating the reliability of obtained PFM images. Accordingly, the microscopic PFM images clearly demonstrate that CG-BSTO undergoes two distinct switching mechanisms: (1) conventional phase switching at shorter time scales, characterized by a discrete reversal of polarization direction (the flip of the PFM phase), and (2) unconventional amplitude switching at longer time scales, in which the polarization magnitude continuously increase while maintaining a fixed direction, as revealed by the increase of PFM amplitude. In addition, the same switching mechanisms of oppositely pre-poled polarization states have been confirmed in BTO and CG-BSTO through PFM measurements, further demonstrating their robustness (Fig. S7).

**Flattened energy minimum in compositionally graded ferroelectric heterostructure revealed by Landau-Ginzburg-Devonshire simulation**

As schematically shown in Fig. 1B, multiple stable states with continuous polarization and consequently the experimentally observed ferroelectric amplitude switching, are only possible when two minimum energy points in the double-well potential becomes flattened. We performed simulations based on the Landau-Ginzburg-Devonshire (LGD) phenomenological theory following the previous studies (*31*) to demonstrate how compositional gradient in CG-BSTO results in the flattening of the mesoscopic energy potential (Material and Method 7). In this framework, the total free energy of the compositionally graded ferroelectric system is expressed as a polynomial function of polarization with each term weighted by the dielectric stiffness coefficients (*31-33*). To estimate the free energy of the homogeneous single $Ba_{1-x}Sr_xTiO_3$ layer with different $Sr^{2+}$ doping ratio, we employed a linear summation approach, where the Landau coefficients of $BaTiO_3$ and $SrTiO_3$ were weighted according to their respective fractions (Fig. 3A). As the $Sr^{2+}$ ion is increased within the homogeneous single $Ba_{1-x}Sr_xTiO_3$ layer, the resulting free energy evolves from ferroelectric to paraelectric landscape. In addition to the layer-by-layer free energies, we included a gradient energy term, $g_{33}(\nabla P)^2$, in the total free energy of the CG-BSTO system to account for spatial variations in polarization. This gradient energy term plays a crucial role in compositionally graded ferroelectrics, where differences in local polarization across layers induce volume bound charges and internal electric fields that destabilize the system energetically. The $g_{33}(\nabla P)^2$ term penalizes abrupt polarization changes and thus captures the energy cost associated with maintaining spatially varying polarization profiles. Note that we omitted the flexoelectric term—which should exist in CG-BSTO—because the anomalous amplitude switching behavior is observed regardless of the polarization direction (Fig. S7). This indicates that the flexoelectric effect plays only a minor role in the switching behavior. The detailed coefficients of LGD equations are given in Table S1.

The effective LGD model simulation for CG-BSTO heterostructure could indeed reproduce pronounced flattening of the energy minimum with small gradient energy term. Based on TOF-SIMS data (Fig. 1C), we constructed an effective LGD model composed of 21 layers where Sr doping ratio linearly changes from 0 to 0.3. By fixing all dielectric stiffness coefficients of each layer listed in Table S1, we performed how the total energy potential varies with respect to gradient parameter $g_{33}$ as shown in Fig. 3B. With large enough $g_{33}$, the total energy potential of CG BSTO becomes similar to the potential of uniform $Ba_{0.85}Sr_{0.15}TiO_3$ (Gray dashed line). This behavior is expected, as a sufficiently large gradient energy term imposes a high energy cost on the spatial variations in polarization and thus strongly suppresses them, effectively enforcing uniform polarization across the heterostructure. Consequently, the total energy is given by a simple linear summation of the energy potentials of individual $Ba_{1-x}Sr_xTiO_3$ layers. As the gradient parameter $g_{33}$ decreases—allowing polarization to vary across layers—the typical double-well potential gradually transforms into an energy landscape with significantly flattened two minima as a function of the mesoscopic ferroelectric polarization (i.e., the average polarization across layers). The flattened energy landscape we found in heterogeneous ferroelectrics aligns well with recent strategies of flattening the free-energy landscape to enhance piezoelectric responses by introducing local heterogeneity in ferroelectrics (*34-36*). Figure 3C illustrates the spatial distribution of local polarization corresponding to specific mesoscopic states near the flattened region of the energy profile. It illustrates how non-uniform local polarization result in continuous mesoscopic polarizations whose energies are almost identical to each other.

Our LGD model with non-uniform microscopic polarization is further supported by previous TEM studies on gradient BSTO heterostructures (*37*). Such polarization discontinuities are generally considered energetically unfavorable due to the formation of volume bound charges, particularly in insulating ferroelectrics that lack free carriers to screen the internal field. However, previous study by Damodaran *et al*. demonstrated that CG BSTO generally possesses sufficient defect density to locally screen uncompensated charges (*37*). This screening suppresses electrostatic energy penalties, allowing substantial polarization gradients to persist, thereby supporting the physical validity of our solution and the resulting flattened free energy landscape.

**Amplitude-switching-induced continuous polarization states from a flattened energy minimum**

Finally, we demonstrate the control of the amplitude switching for analog memory applications. To systematically manipulate the continuous memory states, we applied two distinct voltage pulse schemes: (i) a series of negative pulses (-1.5 V, 20 ms) along the pre-poled direction (Fig. 4A) and (ii) a series of small positive pulses (0.25 V, 100 ms) in the opposite direction (Fig. 4B) (Material and Method 5). In the latter, an electric field smaller than the coercive field was used to prevent the unwanted phase switching. In the first case, successive pulses led to a continuous increase in the normalized switched polarization, $\Delta P(t)/2P_r$, accompanied by a corresponding enhancement in PFM amplitude (Fig. 4C). Conversely, when pulses were applied against the pre-poled state, $\Delta P(t)/2P_r$ gradually decreased, with the PFM amplitude reflecting this trend (Fig. 4D). Notably, throughout both processes, the PFM phase remained unchanged, confirming that the observed polarization-magnitude modulation arises from controlled amplitude switching rather than the phase switching. This controlled amplitude switching is further corroborated by measurements performed under the opposite polarization direction (fig. S8), demonstrating consistent amplitude switching behavior, which arises from the intrinsic symmetry of the LGD potential with two flattened minima. These findings highlight the capability of CG-BSTO heterostructures to achieve continuous polarization states, overcoming the binary limitations of conventional ferroelectric storage and establishing a promising platform for advanced analog memory applications. (*38-41*).

**Discussion**

In the first sight, it may not be trivial that each amplitude-switched polarizations observed in Fig. 4 appears to be nonvolatile despite the LGD potential with flattened minima. One possible mechanism likely involves motion of charged defects, such as oxygen vacancies (*37*). As mentioned above, the presence and redistribution of charged defects is essential to screen the discontinuity of non-uniform polarization and stabilize flattened energy landscapes in CG-BSTO, as shown by the negative bulk charge density arising from non-uniform polarization solution in the LGD simulations (fig. S9). Although each polarization state has nearly similar energy, migration of charged defects requires overcoming an activation barrier, thereby stabilizing the polarization states even in flattened minima (*42*). This defect-mediated mechanism also provides a plausible explanation for the slow dynamics observed in amplitude switching. As illustrated in fig. S10, the time derivative of $P(t)$ reveals two distinct characteristic times: a fast switching on the order of several of microseconds, attributed to the phase switching typical of epitaxial ferroelectrics (*21, 43*). and a much slower process on the order of tens of milliseconds, linked to amplitude switching. The slow dynamics might be governed by the migration of charged defects like oxygen vacancies (*29, 44*).

Finally, we address other possible origins such as relaxor ferroelectricity and leakage current, both of which fail to account for the observed anomalous switching behaviors. At first glance, the slow dynamics and frequency-dependent ferroelectric response observed in our measurements may resemble relaxor-like behavior. However, our CG-BSTO is fundamentally different from typical relaxor ferroelectrics in the following ways. 1) It exhibits a well-defined, square-like *P-E* hysteresis loop across all measurement frequencies from 1 Hz to 5 kHz (Fig. S4). 2) Polarization relaxation measurements (Fig. S5) confirm the excellent stability of the remnant polarization, which would rapidly decay to zero in a typical relaxor ferroelectric. Another possible extrinsic contribution to polarization is leakage current. However, our CG-BSTO exhibits substantially lower leakage current compared to our uniform BTO (Fig. S3), effectively ruling out this possibility as well.

In summary, we demonstrate the existence of energy landscapes with flattened minima in the LGD potential and the amplitude switching of order parameters in compositionally graded ferroelectrics. While the amplitude switching and the LGD potential with flattened minima may seem impossible at microscopic length scales, non-uniform polarization distributions enable these phenomena at mesoscopic length scales (~100 nm). Our findings resonate with the fundamental philosophy of condensed matter physics, which focuses on not the microscopic but the effective Lagrangian in the long-wavelength limit. We believe this approach could extend to other broken-symmetry phases, particularly to continuous symmetry breaking with massless Goldstone modes, potentially leading to even richer physical phenomena.

Moreover, as demonstrated in Fig. 4, the observed amplitude switching can be harnessed as a critical component for next-generation analog memory devices. According to the Landau-Khalatnikov equation (*45*), which describes the time dynamics of order parameters, the LGD potential with flattened minima implies a linear response of polarization to an external electric field. Indeed, our sequential pulse measurement in Fig. 4 exhibits a much more linear response to the number of identical electric pulses than those with conventional ferroelectrics (*38, 39, 41*). Such linearity is crucial for analog memory devices (*46*), and future studies should focus on engineering polarization gradients to enhance linearity, thereby advancing the feasibility of the amplitude switching for analog memory applications.

**Acknowledgments:**

We thank Prof. Jung Hoon Han for helpful discussions on the theoretical interpretation of the results.

**Funding:**

This work was mainly supported by the National Research Foundation (NRF) of Korea funded by the Ministry of Science and ICT (Grant No. NRF-2017M3D1A1040834, RS-2024-00348920, RS-2025-02272971, RS-2025-00512822, RS-2023-NR076385), under the ITRC(Information Technology Research Center) support program(IITP-2024-RS-2023-

00259676). T.H.K. was also supported by the Korea Institute of Science and Technology (KIST) (2E33811).

**Author contributions:**

G.-H.K., T.H.J., S.S., S.M.Y. and C.S. conceptualized this work. G.-H.K., S.S., J.H., J.H.C and H.L. synthesized the thin films and fabricate the device. G.-H.K., S.S., J.H., and H.L. characterized the structural properties of the thin films. T.H.K., and Y.S.O. provided a polycrystalline target for pulsed laser deposition and ferroelectric tester for measuring ferroelectric properties. G.-H.K., T.H.J., S.S., S.C.C., J.K.L and J.H. performed *P-V*, *I-V*, *C-V* measurement, piezoresponse force microscopy, and pulse switching measurement under the direction of S.M.Y. G.-H.K., S.Y.P , P. K. K. and C.S. conducted Ginzburg-Landau-Devonshire simulation. G.-H.K. and J.H. performed scanning tunneling electron microscopy. G.-H.K., T.H.J., S.S., S.M.Y. and C.S. wrote the paper with input from all co-authors. Also,

**Competing interests:**

C.S., S.M.Y., T.H.J., G.-H.K., J.H, and S.S. are inventors on KR patent application(10-2025-0044083) submitted by the Ulsan National Institute of Science and Technology (UNIST) that covers continuous switching memory device using the compositionally graded ferroelectric heterostructure.

**Data and materials availability:** All data are available in the main text or the supplementary materials.

**Supplementary Materials**

Materials and Methods

Figs. S1 to S10

Tables S1

References

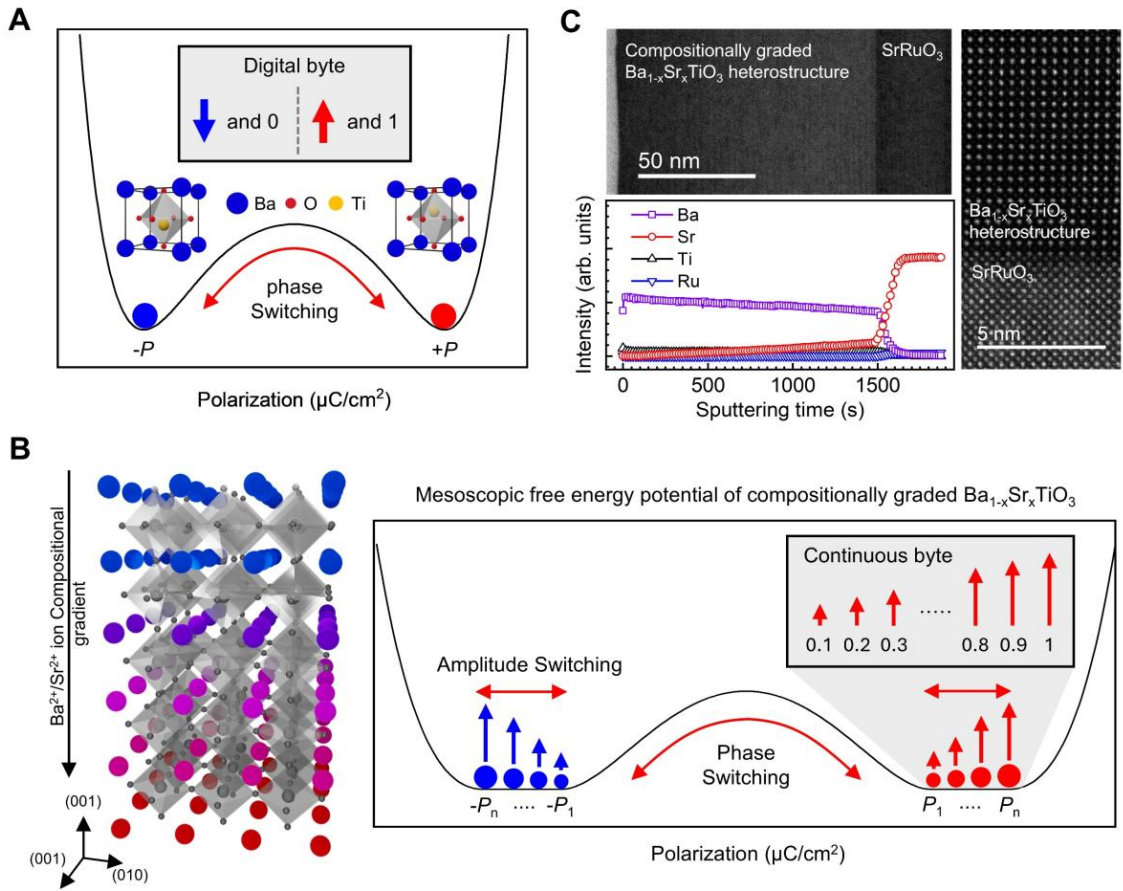

**Fig. 1. Ferroelectric amplitude switching via flattened free energy potential and its application as continuous byte.** (A) Schematic illustration of phase switching in ferroelectric $BaTiO_3$. The polarization state of $BaTiO_3$ is stabilized in two distinct ground states, characterized by upward and downward shifts of the $Ti^{4+}$ ion, corresponding to binary states 0 and 1. (B) Schematic of the compositionally graded $Ba_{1-x}Sr_xTiO_3$ heterostructure and its switching mechanism. The presence of a polarization gradient along the out-of-plane direction alters the free energy landscape, flattening the potential well and enabling polarization amplitude modulation without phase switching, referred to as amplitude switching. This amplitude switching mode allows for continuous polarization states, facilitating continuous byte representation. (C) STEM images and TOF-SIMS analysis of the $Ba_{1-x}Sr_xTiO_3$ heterostructure. The STEM image (top) shows the overall layer structure. ToF-SIMS analysis (bottom) with similar length scale compared to STEM image (top) illustrates the $Ba^{2+}/Sr^{2+}$ compositional gradient, showing a progressive increase/decrease in $Sr^{2+}/Ba^{2+}$ ion concentration from the top surface to the bottom electrode. The high-resolution STEM image (right) reveals the atomic-scale structure of the heterointerface, confirming the epitaxial relationship between $Ba_{1-x}Sr_xTiO_3$ and the $SrRuO_3$ bottom electrode.

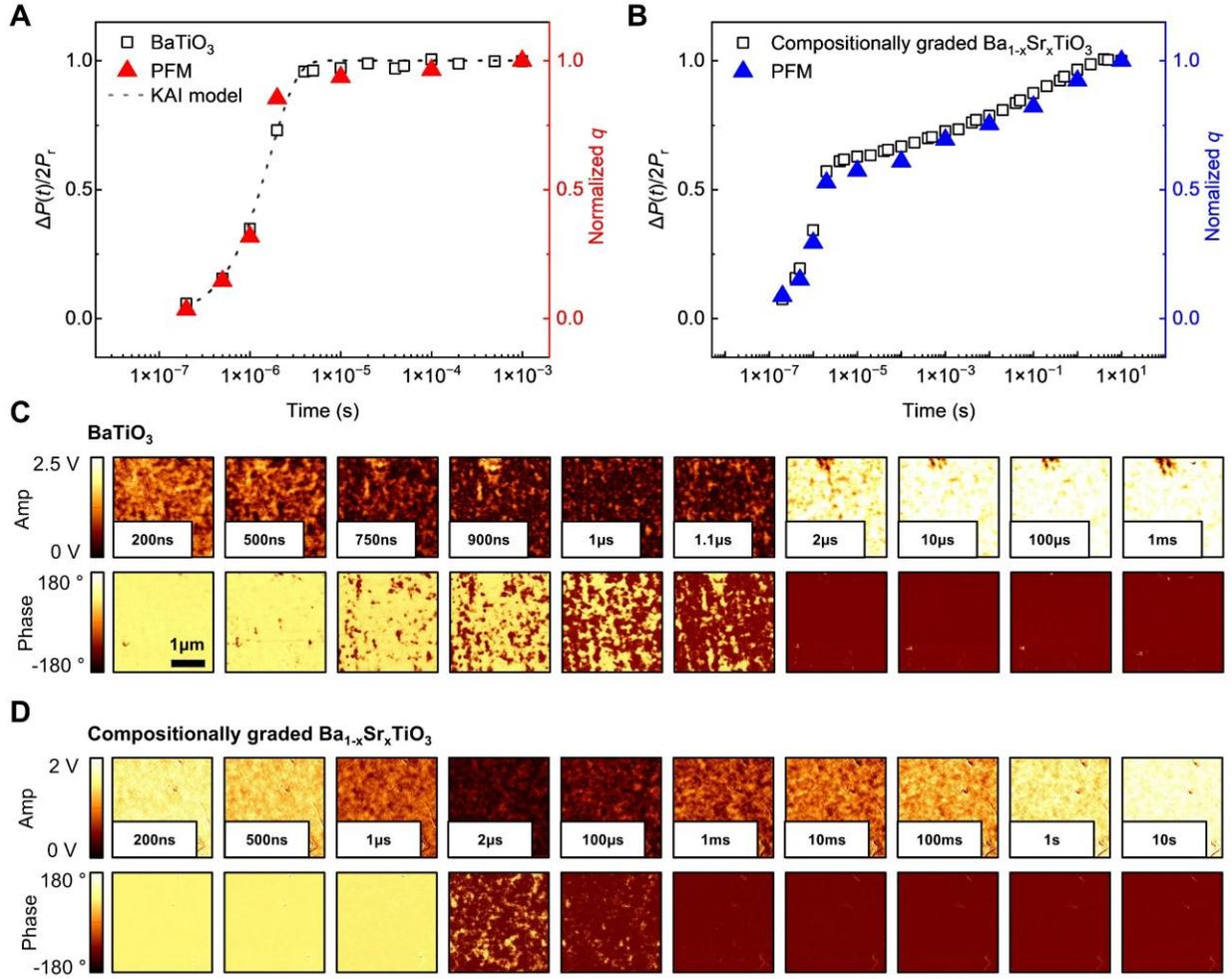

**Fig. 2. Amplitude switching of the compositionally graded Ba$_{1-x}$Sr$_x$TiO$_3$ heterostructure.** Time-dependent $\Delta P(t)/2P_r$ data obtained from switching current measurement (open squares) of (**A**) BTO with the Kolmogorov-Avrami-Ishibashi model fitting (gray dash line) and (**B**) the CG-BSTO heterostructure. Time-dependent PFM images with 3 x 3 μm$^2$ size showing the ferroelectric domain evolution at $V_{ext}$ = 1.5 V of the (**C**) BTO and (**D**) CG-BSTO heterostructure. The amplitude (up) and phase (down) of BTO and CG BSTO are presented separately, and the measure time is written in white box in the images. The normalized switched polarization $\Delta P(t)/2P_r$ of (A) and (B) shows excellent agreements with normalized $q$ value plot, determined from the PFM images (solid triangle), guaranteeing the reliability of the microscopic PFM images.

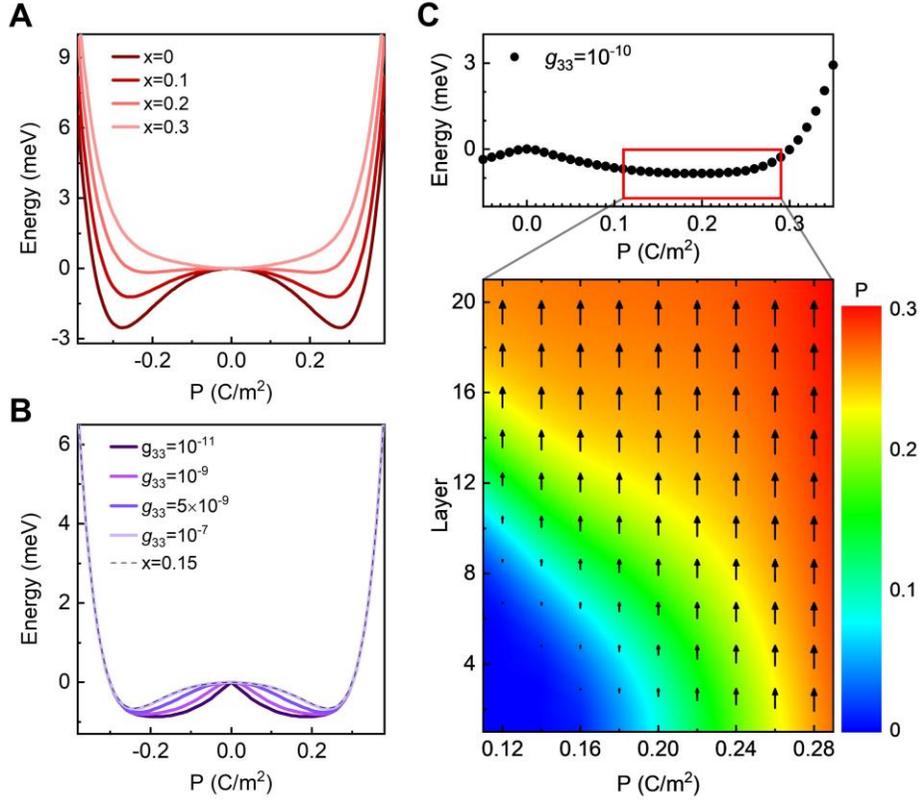

**Fig. 3. The origin of amplitude switching in the compositionally graded Ba$_{1-x}$Sr$_x$TiO$_3$.** (**A**) Landau-Ginzburg free energy potential for homogeneous single Ba$_{1-x}$Sr$_x$TiO$_3$ film with respect to Sr$^{2+}$ ion composition, $x$. As the Sr$^{2+}$ ion composition increases, the potential wall become decreased and finally disappeared like a paraelectric materials. (**B**) Mesoscopic Ginzburg-Landau Free energy potential of compositionally graded Ba$_x$Sr$_{1-x}$TiO$_3$ heterostructure for different gradient energy, $g_{33}$, with fixed compositional gradient (solid lines) and comparison with the energy potential of Ba$_{0.85}$Sr$_{0.15}$TiO$_3$ (a dashed line). (**C**) Local polarization distribution for energy states from $P = 0.12$ (C/m$^2$) to $P = 0.28$ (C/m$^2$) in the free energy potential of $g_{33} = 10^{-10}$. The local polarization gradient exists along the out-of-plane direction of a total of 21 layers that make up the Ba$_{1-x}$Sr$_x$TiO$_3$ heterostructure.

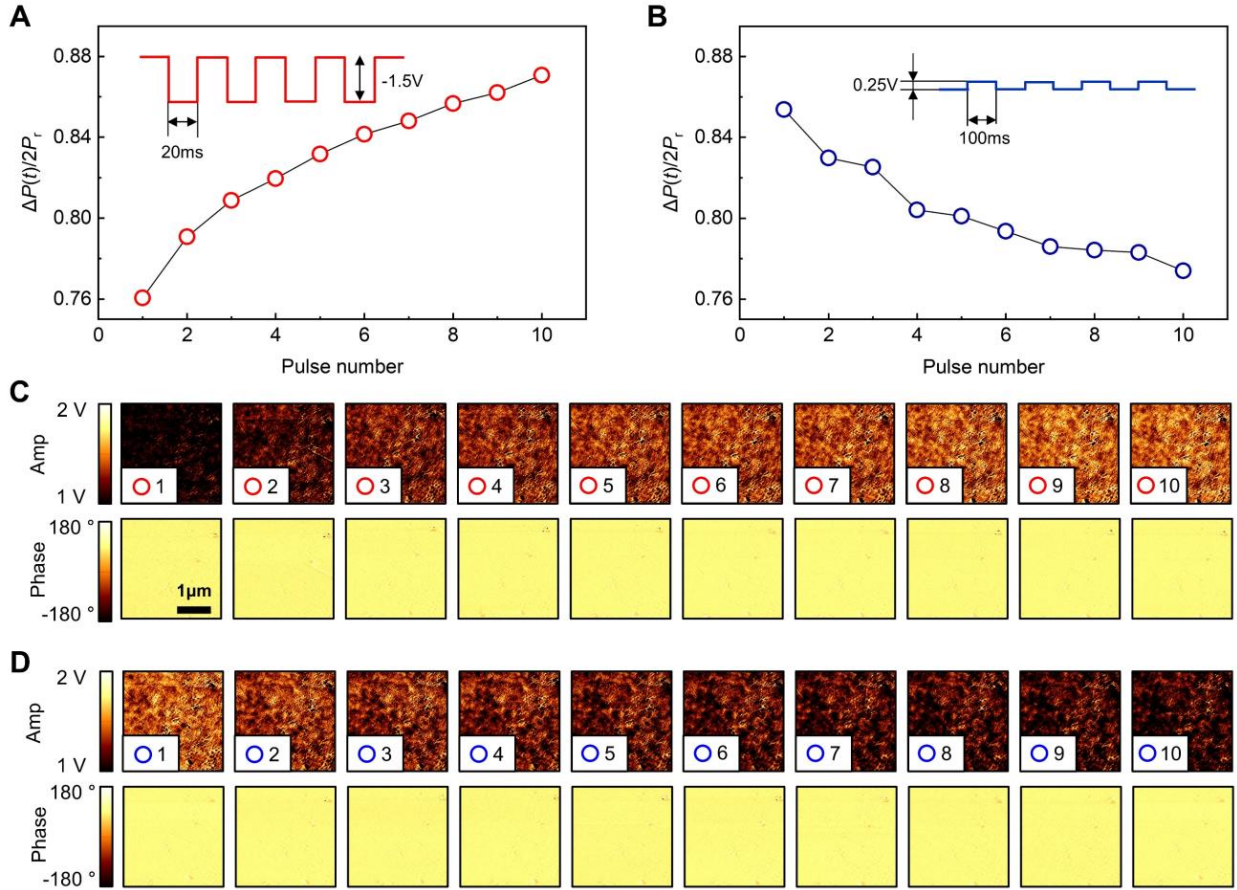

**Fig. 4. The possibility of continuous byte representation in the compositionally graded Ba$_{1-x}$Sr$_x$TiO$_3$ heterostructure.** (**A**) Switching current measurements under a sequence of voltage pulses (–1.5 V, 20 ms) applied in the same direction as the initial polarization, starting from a pre-poled state in the amplitude switching regime. (**B**) Switching current measurements for a pulse applied in the opposite direction of the pre-aligned polarization (0.25 V, 100 ms). The normalized switching polarization, $\Delta P(t)/2P_r$, exhibits a continuous increase or decrease in amplitude, following the respective pulse profiles of (**A**) and (**B**). (**C**) PFM images corresponding to the pulse profile in (**A**), demonstrating a gradual increase in the PFM amplitude while maintaining a stable PFM phase. (**D**) PFM images for the pulse profile in (**B**), showing a continuous decrease in amplitude with the stable PFM phase.

# Supplementary Materials for

## Ferroelectric amplitude switching and continuous memory

Gye-Hyeon Kim [1]†, Tae Hyun Jung[2]†, Seungjoon Sun[1]†, Jung Kyu Lee[2], Jaewoo Han[1], P. Karuna Kumari[3], Jin-Hyun Choi[1], Hansol Lee[1], Tae Heon Kim[3], Yoon Seok Oh[1], Seung Chul Chae[4], Se Young Park[5], Sang Mo Yang[2]*, Changhee Sohn[1]*

Corresponding author: Sang Mo Yang, smyang@sogang.ac.kr and Changhee Sohn, chsohn@unist.ac.kr

**The PDF file includes:**

Materials and Methods
Figs. S1 to S10
Tables S1
References

**Materials and Methods**

1. Sample preparation

Metal–insulator–metal (MIM) capacitors based on $SrRuO_3$ (SRO)/$BaTiO_3$ (BTO)/SRO, SRO/$Ba_{0.9}Sr_{0.1}TiO_3$ (uniform BSTO)/SRO and SRO/compositionally graded $Ba_{1-x}Sr_xTiO_3$ (CG-BSTO)/SRO heterostructures were syntesized using pulsed laser deposition. All films were grown on (110)-oriented $GdScO_3$ single-crystal substrates. The optimized deposition conditions were: a substrate temperature of 750 °C, an oxygen partial pressure of 50 mTorr, and a KrF excimer laser ($\lambda = 248$ nm) with an energy fluence of 1.5 /1.8 J/cm² for SRO/BTO, uniform BSTO and CG-BSTO. The laser repetition rate was set to 5 Hz for the deposition of BTO, uniform BSTO and CG-BSTO layers, and 10 Hz for the SRO electrodes. For the fabrication of uniform BSTO the compositionally graded BSTO layer, layer-by-layer doping was achieved by co-ablating $SrTiO_3$ (STO) and BTO targets. The doping ratio was controlled based on the relative growth rates of STO and BTO thin films, allowing for a estimation of the resulting composition. The total unit cells is designed to be 21 layers for each $BaTiO_3$ thin films with graded Sr doping (~ 100nm), which the one layer possess 12 unit cells.

For fabricating the MIM capacitor structure of SRO/BTO/SRO, SRO/uniform BSTO/SRO and SRO/CG-BSTO/SRO heterostructure, the Photolithographic patterning was carried out using a positive photoresist (GXR 601, 46 cp). The photoresist was spin-coated at 4000 rpm for 30 s, followed by a soft bake at 105 °C for 90 s. UV exposure was performed using a Karl Suss MA6 mask aligner for 10 s. After development, the patterned samples were subjected to wet etching to define the top electrode geometry. For selective removal of SRO, a 0.4 mol/L aqueous solution of sodium periodate ($NaIO_4$), known to effectively etch SRO, was used. The etching rate was approximately 2 nm/s, and an additional 5 s of overetching was applied to ensure complete removal of unwanted SRO residue.

2. Structural/electrical analysis of $BaTiO_3$ and compositionally graded $Ba_{1-x}Sr_xTiO_3$

The crystallographic structure and epitaxial quality of the BTO thin film and the CG-BSTO heterostructure were characterized using high-resolution X-ray diffraction (XRD) and reciprocal space mapping (RSM). All measurements were performed on a Bruker D8 Discover high-resolution X-ray diffractometer equipped with a Cu K-$\alpha_1$ source ($\lambda = 1.5406$ Å). $\theta$–$2\theta$ scans were carried out using a 0D scintillation detector with a step size of 0.01° and a scan speed of 0.5 s per step to confirm the phase purity and out-of-plane orientation of the films. To further investigate the in-plane and out-of-plane lattice parameters, RSM measurements were conducted around the GdScO$_3$ (103)$_{pc}$ asymmetric reflections.

The electrical properties of the film capacitors were investigated using a ferroelectric tester (TF analyzer 2000E, aixACCT) to measure $P$–$V$ hysteresis loops, capacitance–voltage ($C$–$V$) curves, and $I$–$V$ curves. For the $P$–$V$ hysteresis loops, triangular pulses with a frequency of 1 kHz and an amplitude of 2 V were employed. The $\varepsilon_r$ values were extracted from the $C$–$V$ data, obtained by applying a small AC signal of 1 kHz frequency and 50 mV amplitude.

3. Scanning tunneling electron microscopy

The cross-sectional lamella of the CG-BSTO heterostructure was prepared using a dual-beam focused ion beam system (Helios NanoLab 450, FEI) with a Ga-ion beam operated at an acceleration voltage of 30 kV. Atomic-resolution annular dark-field scanning transmission electron microscopy (ADF-STEM) images were obtained using a Cs-corrected STEM (JEM-ARM200F, JEOL) operated at 160 kV. Elemental mapping was conducted using energy-dispersive X-ray spectroscopy (EDS) in STEM mode to visualize the spatial distribution of Ba and Sr across the CG-BSTO layer. The Ba L-edge and Sr L-edge signals were collected to confirm the compositional gradient, which reveals an increasing Sr content toward the bottom SRO electrode. The resulting elemental maps and Sr concentration profiles confirm successful implementation of the designed gradient.

4. Depth chemical profiling using time-of-flight secondary ion mass spectroscopy (TOF-SIMS)

TOF-SIMS depth profiling of the CG-BSTO heterostructure was performed using a TOF-SIMS V instrument (IONTOF GmbH, Germany) in positive ion mode. A pulsed Bi$^+$ ion beam with an energy of 25 keV and a current of 1.00 pA was used for analysis over a 50 × 50 μm² area. For sputtering, an O$_2^+$ ion beam with an energy of 0.5 keV and a current of 85.0 nA was used over a 250 × 250 μm² area. The measured secondary ion intensities for Ba and Sr were used to construct the compositional depth profile across the film thickness. The results confirm a gradual increase in Sr concentration toward the bottom electrode, consistent with the designed compositional gradient.

5. Switching current measurements

Time-dependent switchable polarization values, $\Delta P(t)$, of the BTO and the CG-BSTO MIM capacitors were obtained through switching current measurements. $\Delta P(t)$ is obtained from the difference between the switched polarization ($P_{sw}$) and non-switched polarization ($P_{ns}$) states, i.e., $\Delta P(t) = P_{sw}(t) - P_{ns}(t)$. The $P_{ns}$ state was determined by integrating the current measured during the application of a reading pulse ($\pm 2$ V, 1 s) with the same amplitude and duration as the writing pulse, following the initial writing step. The $P_{sw}$ state is obtained by integrating the current measured during a reading pulse ($\pm 2$ V, 1 s) applied in the direction opposite to that of the writing pulse, where the writing pulse was applied with an experimentally adjusted amplitude and duration.

Subsequently, to investigate the continuous polarization state in CG-BSTO, $\Delta P(t)$ was calculated using the same method. The $P_{ns}$ state was measured by applying the same pulse conditions as described above. For the $P_{sw}$ state, the initial polarization condition was prepared differently for the continuous increase and decrease measurements. To observe the continuous increase in polarization, a single $\pm 1.5$ V, 20 ms writing pulse was applied to create a partially switched state. In contrast, for the continuous decrease in polarization, the $\pm 1.5$ V, 20 ms pulse was applied repeatedly for 11 cycles to induce a more switched initial state compared to the potentiation condition, while intentionally avoiding full polarization saturation. Subsequently, a series of identical pulses ($\pm 1.5$ V, 20 ms) was applied to induce continuous polarization increase, whereas a series of pulses in the opposite direction ($\mp 0.25$ V, 100 ms) was used to induce continuous polarization decrease. After each sequence, the current was integrated during the application of a reading pulse to quantify the polarization state. Switching current measurements involved the use of a digital oscilloscope (DLM3024, Yokogawa) and arbitrary waveform generators (FG410 and FG420, Yokogawa), all managed by custom LabVIEW software.

6. Piezoresponse force microscopy

For nanoscale analysis of ferroelectric domain switching dynamics on the capacitors top electrode, a commercial atomic force microscope (NX10, Park Systems) equipped with an additional probe needle (APN-PFM) was used (*27*). To minimize artifacts such as electrostatic effects, non-conductive cantilevers (PPP-FMR, Nanosensors) were used. During PFM imaging, an AC bias of 0.2 V—below the coercive voltage—was applied with a near-contact resonance frequency of approximately 330 kHz to the top electrode using the APN. To enhance the signal-to-noise ratio significantly in the PFM images, the dual frequency resonance tracking (DFRT) technique was utilized with a lock-in amplifier (HF2LI, Zurich Instrument) (*30*). The normalized piezoresponse, $q$ value represents by $R\cos\theta$, where the $R$ denotes the PFM amplitude and $\theta$ is the PFM phase obtained from the PFM images. To quantitatively evaluate the change in the $q$ value upon pulse application, the deviation from the pre-poled state was normalized by the total difference between the saturated $q$ values corresponding to the two fully switched polarization states. This normalized quantity is defined as the normalized $q$ value.

7. Landau-Ginzburg-Devonshire (LGD) model simulation

The total LGD free energy $G_{tot}$ for compositionally graded $Ba_{1-x}Sr_xTiO_3$ includes both the bulk free energy $G_{uni}$ and the gradient energy $G_g$. The bulk free energy $G_{uni}$ for each $Ba_{1-x}Sr_xTiO_3$ layer is expressed as an eighth-order polynomial of polarization:

$$G_{uni} = A_1 p^2 + A_{11} p^4 + a_{111} p^6 + a_{1111} p^8 \quad (1)$$

where $p$ is local polarization of each $Ba_{1-x}Sr_xTiO_3$ layer, which is along the [001] direction, $A_1$ $A_{11}$, $a_{111}$, $a_{1111}$ are dielectric stiffness coefficients. Especially, $A_1$ and $A_{11}$ are the modified dielectric stiffness coefficients defined as,

$$A_1 = a_1 - \frac{2Q_{12}}{s_{11} + s_{12}} u_m, \qquad A_{11} = a_{11} + \frac{Q_{12}^2}{s_{11} + s_{12}} \quad (2)$$

where $Q_{12}$ is the cubic electrostriction coefficients, $s_{11}$ and $s_{12}$ are the cubic elastic compliances at constant dielectric displacement. And the gradient energy $G_g$ is written as,

$$G_g = g_{33}(\nabla p)^2 \quad (3)$$

By the ref (*28*) in the main text, the depolarization field effect is negligible in the CG-BSTO system, so that we can ignore the depolarization field term in the total free energy and it can be described as,

$$G_{tot} = \int (G_{uni} + G_{grad}) dz = \int \left[ A_1 p^2 + A_{11} p^4 + a_{111} p^6 + a_{1111} p^8 + g_{33}\left(\frac{dp}{dz}\right)^2 \right] dz \quad (4)$$

Considering our structure geometry of CG-BSTO, we can construct the total free energy of CG-BSTO with 21 individual layers with different local polarization,

$$G_{tot} = \sum_{i=1}^{21} \frac{1}{21}[A_1 p_i^2 + A_{11} p_i^4 + a_{111} p_i^6 + a_{1111} p_i^8] + g_{33}\left(\frac{p_{i+1} - p_i}{d}\right)^2$$

where $p_i$ is the local polarization in the i-th layer, and $d$ is the layer thickness. To obtain the equilibrium polarization profile of minimized total free energy, the total free energy was minimized by numerically using Mathematica (FindMinimum), which relies on Lagrangian-based numerical optimization methods under the constraint $p_i > 0$, with fixed average polarization, $P$ (C/m$^2$). The results were stable with respect to various initial conditions, indicating convergence to a physically meaningful solution.

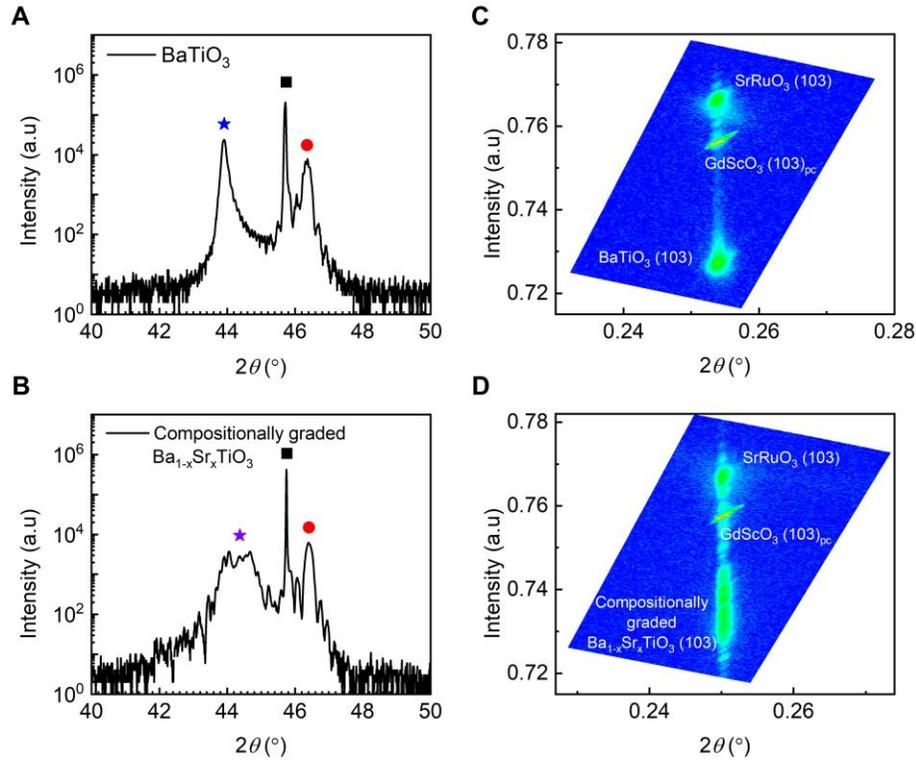

**Fig. S1. Structural analysis by high-resolution x-ray diffraction.** (**A**) XRD $\theta$-$2\theta$ scan of a BTO thin film grown on a GSO (110) substrate. (**B**) XRD $\theta$-$2\theta$ scan of a CG-BSTO heterostructure grown on the same substrate. The GSO (220) peak (blue square), SRO (002) peak (green circle), and BTO (002) peak (red star) are marked. The broad, flat (002) peak corresponding to the (002) peak of CG-BSTO heterostructure is denoted by the purple star. (**C** and **D**) Reciprocal space mapping (RSM) images around the GSO (103)$_{pc}$ reflections for (**C**) the BTO thin film and (**D**) the CG-BSTO heterostructure. The alignment of SRO (103), GSO (103)$_{pc}$, and BTO or CG-BSTO (103) peaks along the same $q_x$ axis indicates coherent in-plane lattice matching without relaxation.

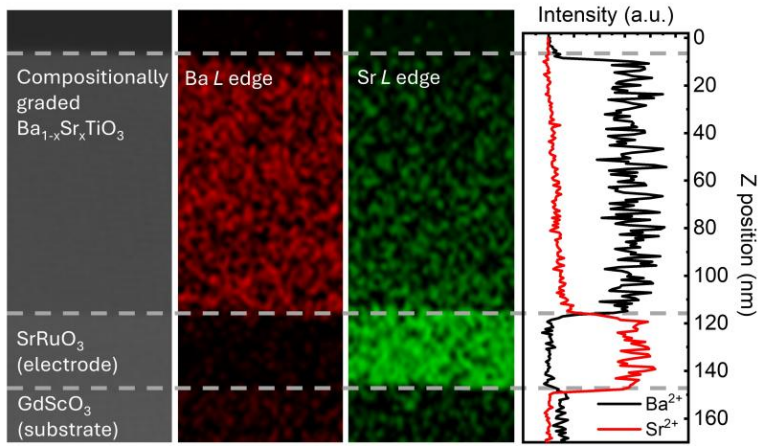

**Fig. S2. Compositional gradient mapping in compositionally graded Ba$_{1-x}$Sr$_x$TiO$_3$ heterostructure.** Cross-sectional annular dark-field scanning transmission electron microscopy (ADF-STEM) image (left), elemental maps for Ba (Ba L-edge, red) and Sr (Sr L-edge, green) acquired by EDS (middle), and the corresponding Sr concentration profile (right) for compositionally graded Ba$_{1-x}$Sr$_x$TiO$_3$ heterostructure. The Sr content increases gradually from the top surface toward SrRuO$_3$ electrode, confirming the intended compositional gradient.

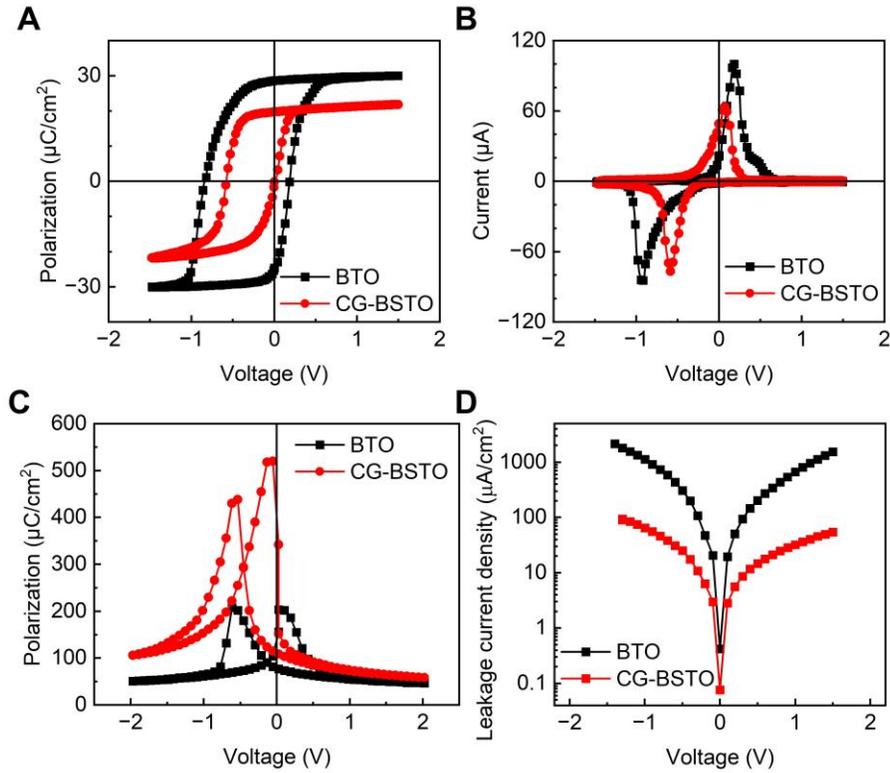

**Fig. S3. Ferroelectric properties in BaTiO$_3$ and compositionally graded Ba$_{1-x}$Sr$_x$TiO$_3$. (A)** *P-V* hysteresis loops of BTO (black square) and the CG-BSTO (red circle). (**B**) corresponding *I-V* loops and (**C**) *C-V* loops of BTO, and CG-BSTO. All measurements confirm ferroelectric properties in both BTO and CG BSTO heterostructure. (**D**) Leakage current density of BTO and CG-BSTO. It is noted that the leakage current of CG-BSTO is the order of magnitude lower than that of BTO.

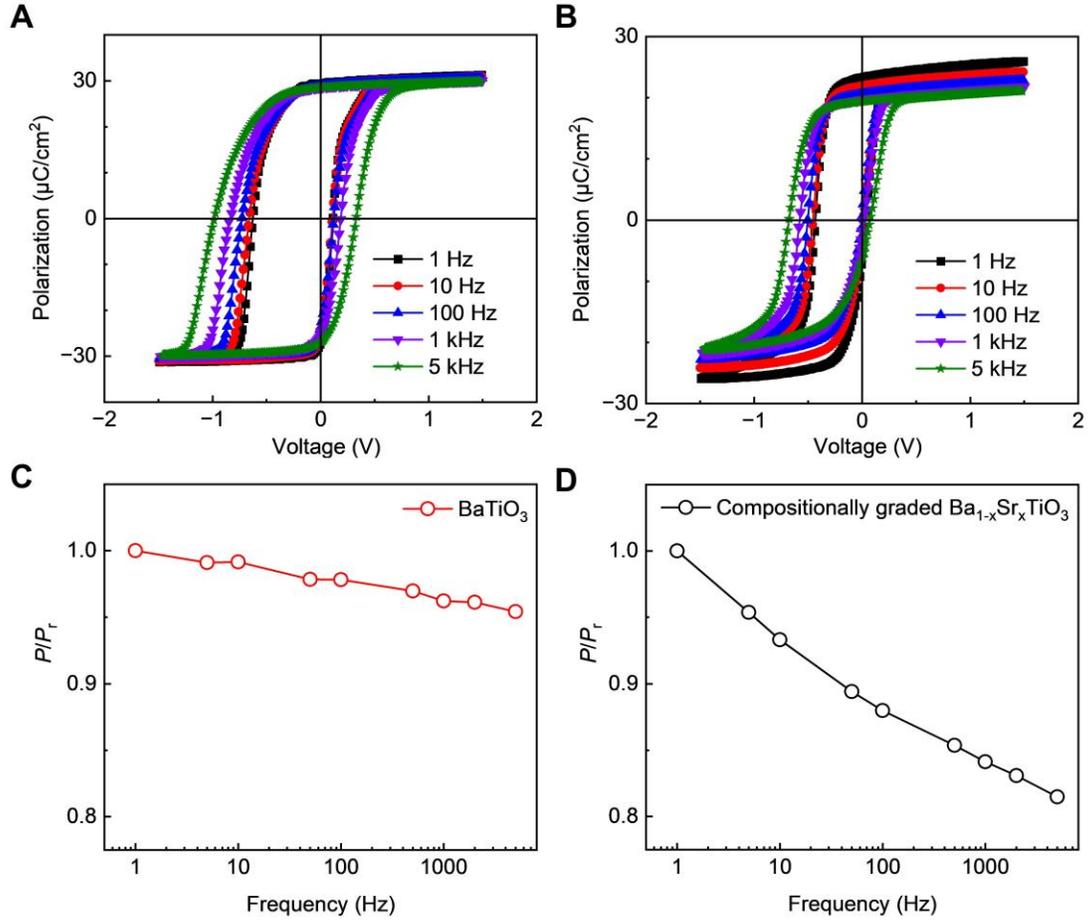

**Fig. S4. Frequency dependence of polarization behavior in BaTiO₃ and compositionally graded Ba₁₋ₓSrₓTiO₃.** (**A**, **B**) *P-V* hysteresis loops measured under bipolar driving at various frequencies (1 - 5kHz) for (**A**) BTO thin film and (**B**) the CG-BSTO heterostructure. Both the *P-V* hysteresis loop of BTO and CG-BSTO shows small differences between saturated polarization ($P_{max}$) and remanent polarization ($P_r$), denoted as $P_{max}$-$P_r$, indicating both systems show typical ferroelectric behavior. As the frequency increases, the coercive voltage increases in both materials, but the decrease in remanent polarization ($P_r$) is more significant in the graded film. (**C**, **D**) Extracted $P_r$ values as a function of frequency for (**C**) the BTO and (**D**) the CG-BSTO. In BTO thin film, $P_r$ remains nearly constant over the measured frequency range. In contrast, the CG-BSTO heterostructure exhibits a strong dependence of $P_r$ on frequency, attributed to the increased time available for more amplitude switching at lower frequencies.

    10

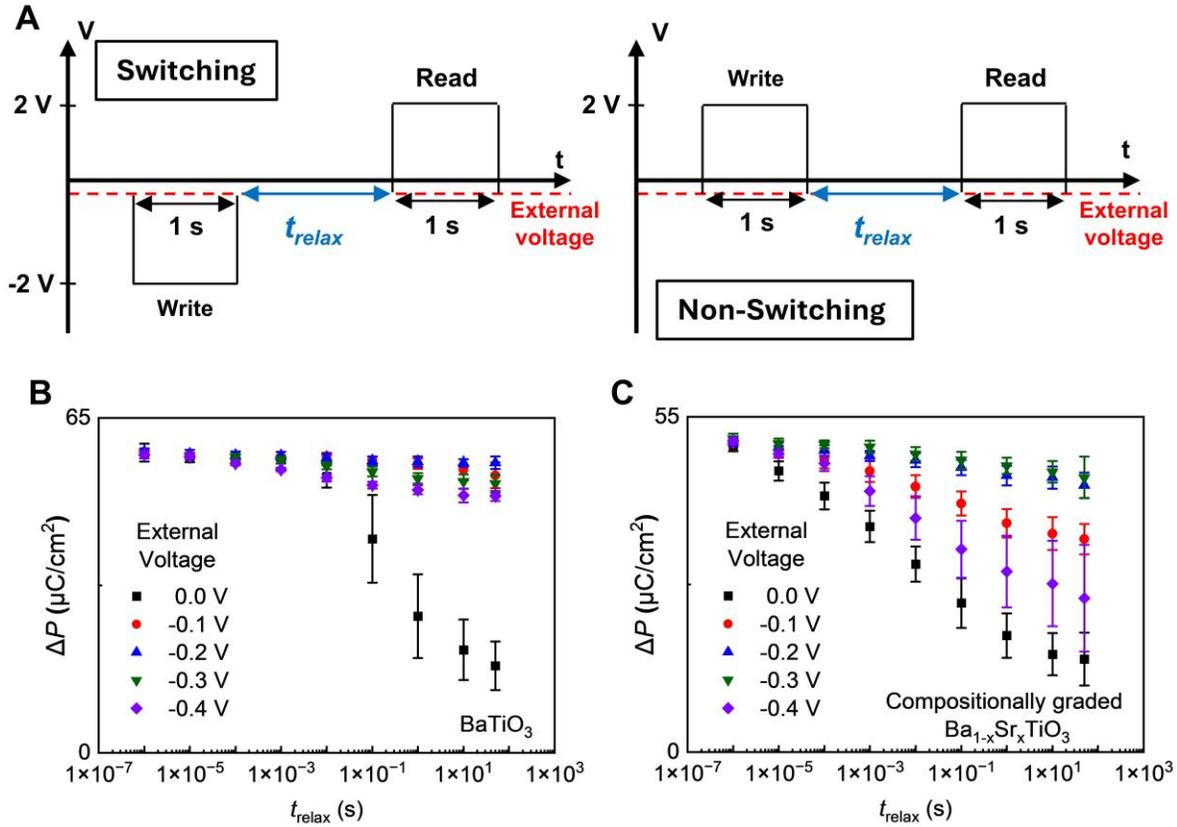

**Fig. S5. Polarization relaxation measurement for BaTiO$_3$ and compositionally graded Ba$_{1-x}$Sr$_x$TiO$_3$.** (**A**) Schematic of the pulse sequence for the polarization relaxation measurements. By subtracting non-switching from switching polarization $\Delta P$ can be obtained. Polarization relaxation result in (**B**) BTO and (**C**) the CG-BSTO by adjusting the $t_{relax}$ and external voltage. Without an external bias, a reduction in polarization was observed in both BTO and CG-BSTO as $t_{relax}$ increased, indicating polarization relaxation behavior. This relaxation behavior is attributed to the imprint effect, which evident from the *P-V* hysteresis loops in Fig. S3 (**A**). When an external voltage was applied during the relaxation period, a significant suppression of polarization decay was observed in both materials, suggesting that the external field can stabilize the polarization state and mitigate the imprint-induced relaxation.

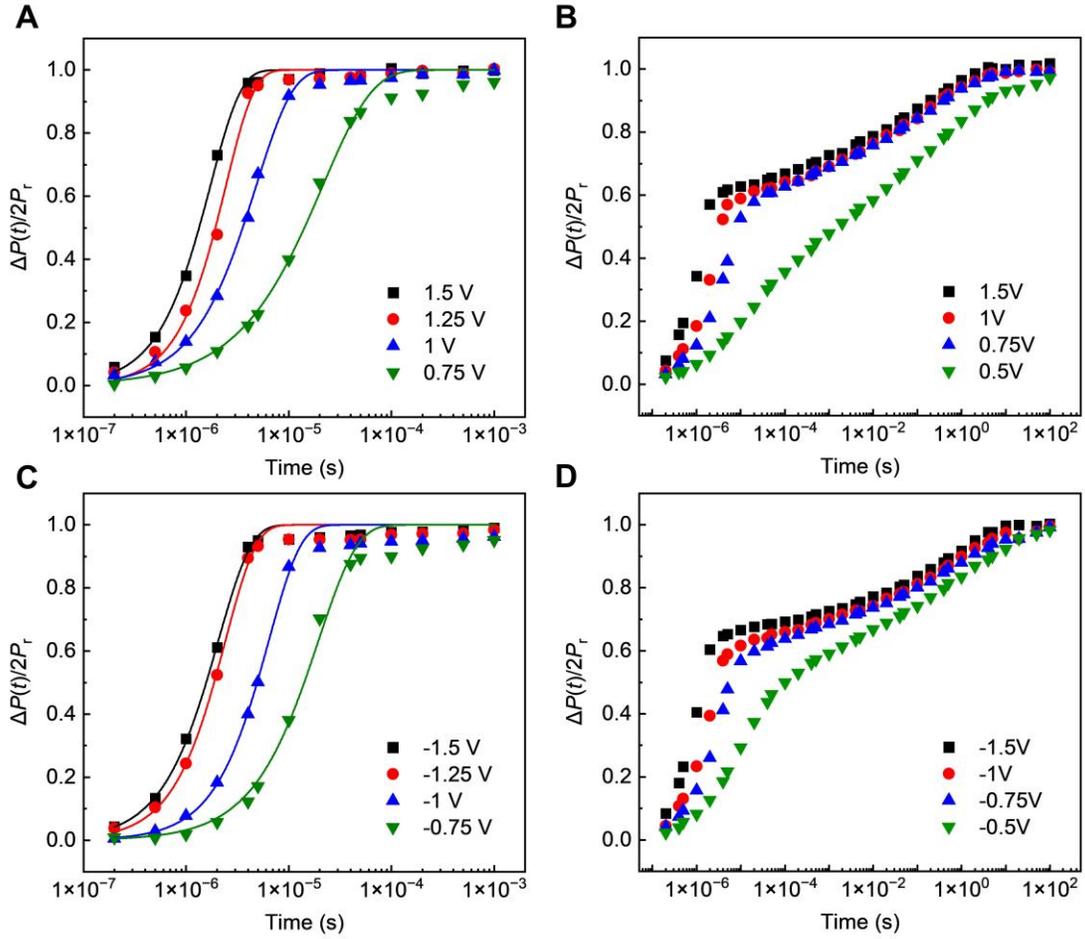

**Fig. S6. Switching current measurement with different voltage pulse.** Time-dependent $\Delta P(t)/2P_r$ data as a function of different writing voltage with -0.2 V external voltage to cancel out the built-in electric field. The positive biases result in (**A**) BTO and (**B**) the CG-BSTO, and the negative biases result in (**C**) BTO and (**D**) the CG-BSTO. While the switching behavior in BTO follows the KAI model, the CG-BSTO heterostructure deviates from the KAI model, suggesting a different switching mechanism originated by compositional gradient. Notably, both BTO and CG-BSTO exhibit nearly identical switching characteristics under all the positive and negative bias conditions.

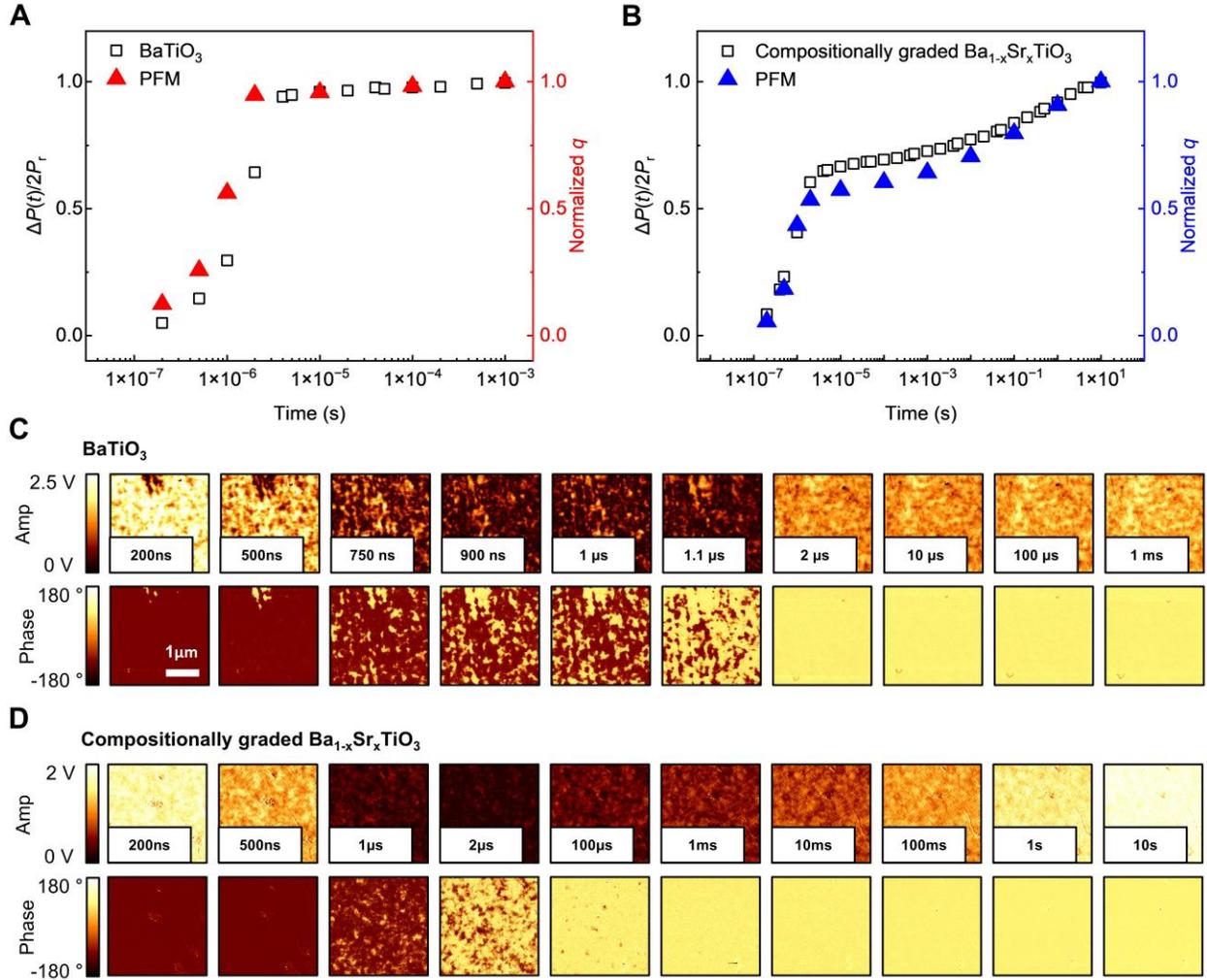

**Fig. S7. Amplitude switching of the compositionally graded $Ba_{1-x}Sr_xTiO_3$ heterostructure with oppositely pre-poled polarization state.** Time-dependent $\Delta P(t)/2P_r$ data obtained from switching current measurement (open squares) and the $q$ values determined from the PFM images (solid triangle) of (**A**) BTO and (**B**) the CG-BSTO heterostructure. Time-dependent PFM images with 3 x 3 μm² size showing the ferroelectric domain evolution at $V_{ext}$ = -1.5 V of the (**C**) BTO and (**D**) the CG-BSTO heterostructure. The amplitude (up) and phase (down) of BTO and CG-BSTO are presented separately, and the measure time is written in white box in the images. The normalized $q$ value shows excellent agreements with the normalized switching polarization $\Delta P(t)/2P_r$ of (**A**) and (**B**), guaranteeing the reliability of the microscopic PFM images.

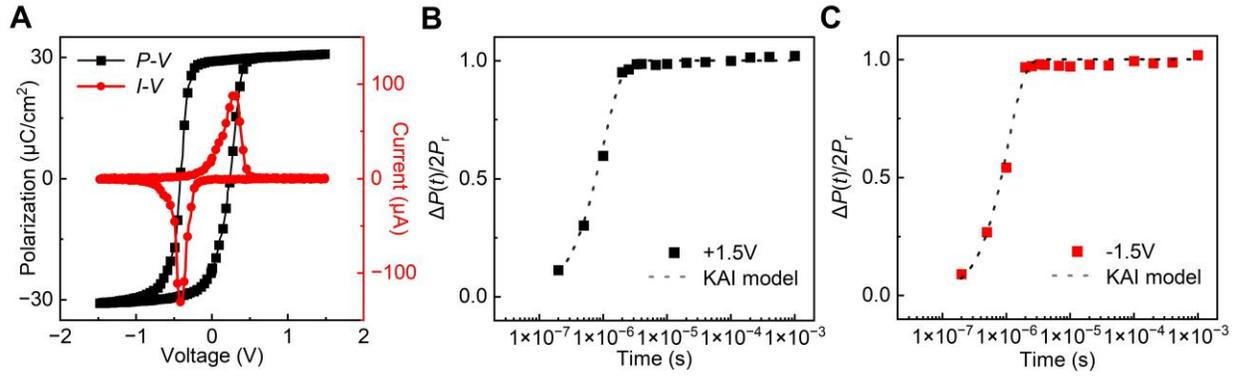

**Fig. S8. Conventional ferroelectric properties in uniformly doped Ba$_{0.9}$Sr$_{0.1}$TiO$_3$ thin film.** (A) *P-V* hysteresis loop (black square) and *I-V* curve (red circle) of uniform BSTO thin film. Time-dependent $\Delta P(t)/2P_r$ data (open squares) of uniform BSTO with the KAI model fitting (gray dash line) with (B) +1.5V pre-poled pulse and (C) -1.5V pre-poled pulse. Uniform BSTO thin film exhibits ferroelectric behavior similar to pure BaTiO$_3$, with well-defined hysteresis characteristic of KAI-type polarization switching. In contrast, uniform doping does not induce the ferroelectric amplitude switching observed in CG-BSTO.

| Parameter | BaTiO$_3$ | SrTiO$_3$ | Compositionally graded Ba$_{1-x}$Sr$_x$TiO$_3$ |
|---|---|---|---|
| $a_1$ (10$^5$ m$^2$N/C$^2$) | $3.61 \times (T - 391)$ | $405 \times \left(\coth\left[\frac{54}{T}\right] - \coth\left[\frac{54}{30}\right]\right)$ | $\left(405\left(\coth\left[\frac{54}{T}\right] - \coth\left[\frac{54}{30}\right]\right) - 3.61(T-391)\right)x + 3.61(T-391)$ |
| $a_{11}$ (10$^8$ m$^6$N/C$^4$) | $-18.3 + 0.04T$ | $1.04$ | $(19.34 - 0.04T)x + (-18.3 + 0.04T)$ |
| $a_{111}$ (10$^9$ m$^{10}$N/C$^6$) | $(13.9 - 0.032T)$ | - | $(13.9 - 0.032T)(1-x)$ |
| $a_{1111}$ (10$^{10}$ m$^{10}$N/C$^6$) | $4.84$ | - | $4.84(1-x)$ |
| $Q_{12}$ (m$^4$/C$^2$) | $-0.034$ | $-0.0135$ | $(0.0209x - 0.034)$ |
| $s_{11}$ (10$^{-12}$ m$^2$/N) | $9.1$ | $3.729$ | $-5.371x + 9.1$ |
| $s_{12}$ (10$^{-12}$ m$^2$/N) | $-3.2$ | $-0.9088$ | $2.291x - 3.2$ |
| $u_m$ | $a_{BTO} = 0.3986$ | $a_{STO} = 0.3905$ | $(-0.0022 - 0.0081x)/(-0.0081x + 0.3986)$ ($\because a_{sub} = 0.3964$) |

**Table S1. Thermodynamic and electromechanical parameters used for modeling BaTiO$_3$, SrTiO$_3$ and compositionally graded Ba$_{1-x}$Sr$_x$TiO$_3$.** Dielectric stiffness coefficients ($a_1$, $a_{11}$, $a_{111}$, $a_{1111}$), electrostrictive coefficient ($Q_{12}$), elastic compliances ($s_{11}$, $s_{12}$), misfit strain ($u_m$) defined as ($a_{sub}$-$a_{film}$/$a_{film}$) for compositionally graded Ba$_{1-x}$Sr$_x$TiO$_3$ heterostructure in given temperature, $T$ and Sr$^{2+}$ ion concentration, $x$ (*31-33*). Parameters for the graded composition are derived by interpolation or analytical combination of BaTiO$_3$ and SrTiO$_3$ values at the room temperature ($T$=300 K).

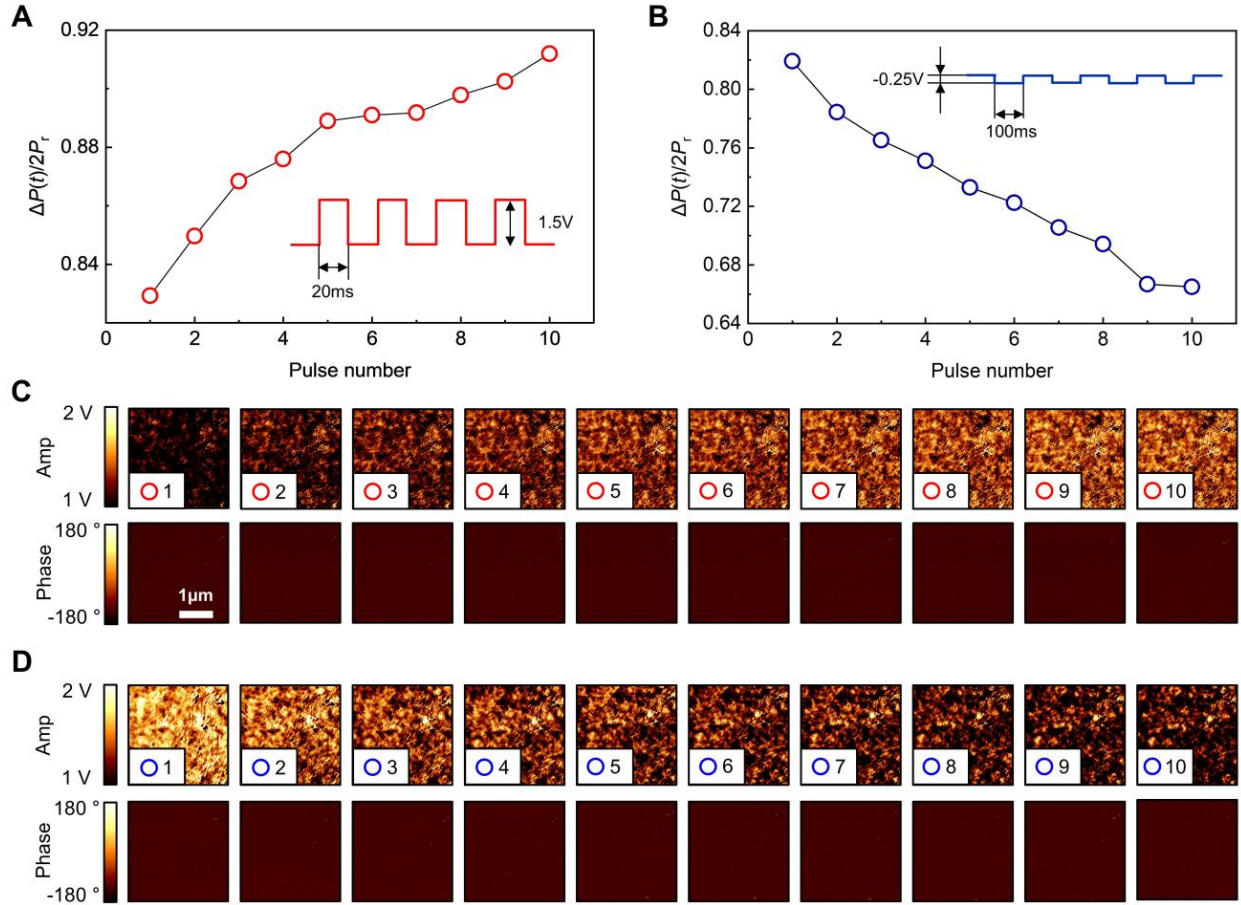

**Fig. S9. Continuous polarization states in opposite pre-poled compositionally graded Ba$_{1-x}$Sr$_x$TiO$_3$ heterostructure.** (**A**) Switching current profiles under a series of pulses (1.5 V, 20 ms) applied to the initially oppositely pre-poled state, compared to the case of Fig. 4A of the main text. (**B**) Switching current profiles under a series of opposite pulses (-0.25 V, 100 ms) applied to the poled state, demonstrating a progressive reduction in the polarization amplitude. (**C**) PFM amplitude and phase images corresponding to (**A**), showing a continuous increase in amplitude with stable phase contrast over 10 successive pulses. (**D**) PFM amplitude and phase images for (**B**), illustrating a systematic decrease in polarization amplitude with preserved phase stability. These results, obtained under reversed bias conditions, confirm the symmetric and robust ferroelectric amplitude switching, further supporting the feasibility of continuous analog memory operation.

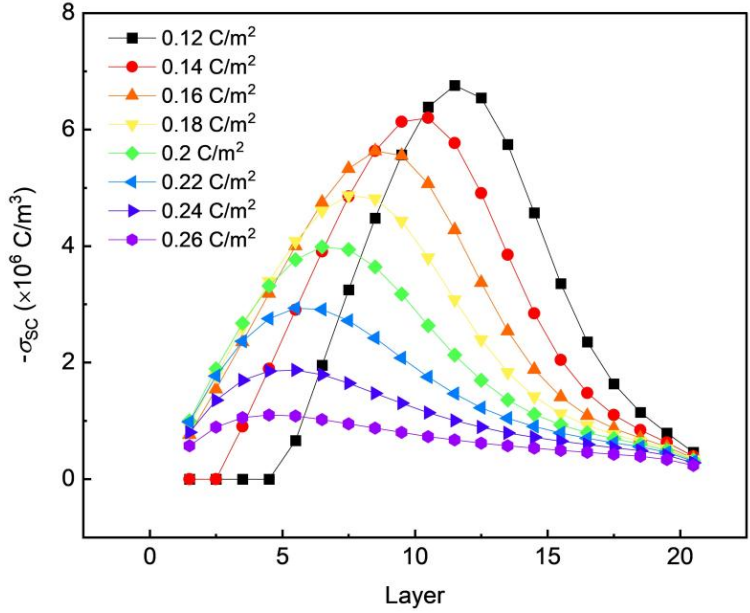

**Fig. S10. Negative charge distribution from non-uniform microscopic polarizations and its evolution with respect to each mesoscopic (average) polarization.** Simulated layer-by-layer distribution of charge density ($-\sigma_{sc}$) across the heterostructure, calculated from the gradient of polarization profiles obtained via Landau-Ginzburg-Devonshire modeling (Fig. 3C in the main manuscript). Each curve represents a different mesoscopic polarization state, labeled by their corresponding average polarization magnitude. As the average polarization increases, the charge distribution exhibits a pronounced shift toward the bottom interface.

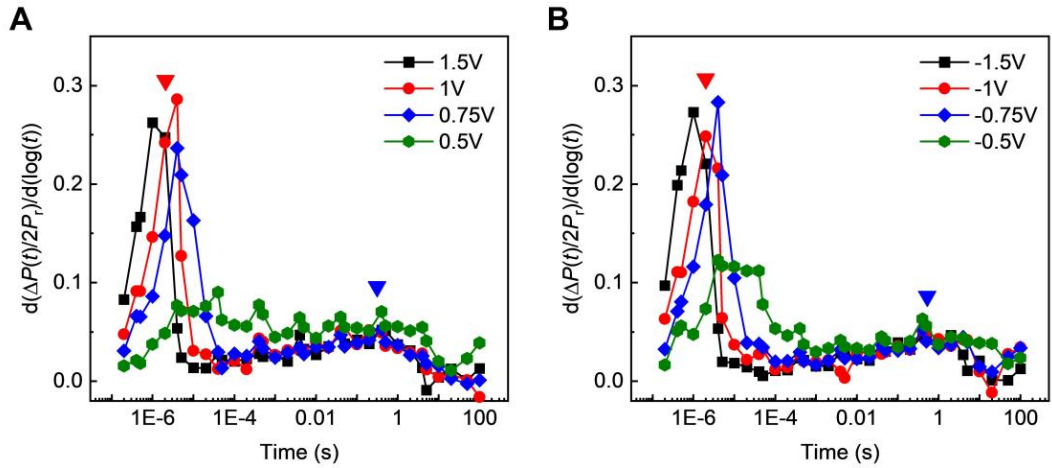

**Fig. S11. Two characteristic times in the switching dynamics of compositionally graded Ba$_{1-x}$Sr$_x$TiO$_3$ heterostructure.** Characteristic switching times extracted by taking the derivative of the switching current with respect to log(time) for (**A**) positive bias and (**B**) negative bias. Two distinct regimes are identified: In phase switching regime (red triangle), the characteristic time decreases with increasing voltage, consistent with field-driven domain wall motion. In contrast, the amplitude switching regime (blue triangle) exhibits a slow and voltage-independent characteristic time.